\documentstyle[twoside,fleqn,espcrc2]{article}

\newcommand{\AmS}{{\protect\the\textfont2
  A\kern-.1667em\lower.5ex\hbox{M}\kern-.125emS}}

\title{Architectural choices for the Columbia 0.8 Teraflops machine}

\author{I. V. Arsenin\address{Physics Department,
	Columbia University,\\
        New York, NY 10027, U.S.A.}%
        \thanks{Work done in collaboration with D.~Chen, N.~Christ, 
C.~Jung, A.~Kahler,  Y.~Luo, 
R.~Mawhinney,  P.~Vranas at Columbia Univ; A.~Gara, J.~Parsons at Nevis
Labs; R.~Edwards, A.~Kennedy at SCRI; S.~Hansen at Fermilab; J.~Sexton
at Trinity College, Dublin; G. Kilcup at Ohio State Univ.  Research is 
sponsored in part by the Department of Energy. } 
}
\begin{document}
  \def\thepage{CU--TP--662 \ \ \ hep-lat/9412093}
  \thispagestyle{myheadings}

\begin{abstract}
	We discuss the hardware design choices made 
	in our $16K$-node 0.8 Teraflops supercomputer project, a machine 
	architecture optimized for full QCD calculations.
	The efficiency of the conjugate gradient algorithm in terms of balance
	of floating-point operations, memory handling and utilization,
	and communication overhead is addressed. 
	We also discuss the technological innovations and
	software tools  that facilitate hardware design and 
	what opportunities 
	these give to the academic community.
\end{abstract}

\maketitle

\section{OVERVIEW}

The preliminary discussions of the Columbia 0.8 Teraflops project began
in Spring 1993. Since 1989 the Columbia group has performed lattice QCD
calculations on the Columbia 256-node parallel computer with 16 Gflops 
peak speed. It  became
clear by that time that the capacity of this computer will have been
exhausted in the course of few years and it was necessary to look into
ways to enhance the computing power in order to stay at the
forefront of lattice calculations.

The timing of our decision to embark on the building of a new supercomputer
was greatly influenced by advances in technology that have produced inexpensive
and relatively fast processors, external memory units, and peripherials
that could be customized for the desired tasks. Our choice was in favor
of a machine that consists of a large number of simple  
nodes connected by an efficient communication network --- a
configuration, we believe, more promising and cost efficient than a few
high speed and high capacity processors which have yet to become 
available at reasonable prices. 

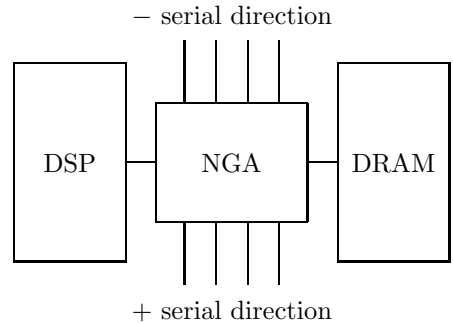
\begin{figure}[htb]
\setlength{\unitlength}{1.05mm}
\begin{picture}(55,45)(-7,0)

\put(0,10){\framebox(14,25){DSP}}
\put(41,10){\framebox(14,25){DRAM}}
\put(18,15){\framebox(19,15){NGA}}
\multiput(21.5,30)(4,0){4}{\line(0,1){8}}
\multiput(21.5,15)(4,0){4}{\line(0,-1){8}}
\put(14,22.5){\line(1,0){4}}
\put(37,22.5){\line(1,0){4}}
\put(27.5,5){\makebox(0,0)[t]{$+$ serial direction}}
\put(27.5,40){\makebox(0,0)[b]{$-$ serial direction}}

\end{picture}

\caption{The functional diagram of one of the $16K$ nodes.}
\label{fig:node}
\end{figure}

\begin{figure}[htb]
\setlength{\unitlength}{0.9mm}
\begin{picture}(75,50)(-4,0)

\put(12.5,35){\framebox(50,15){\shortstack{DSP I/O \\ controller}}}
\put(0,20){\framebox(25,10){SCU}}
\put(45,20){\framebox(30,10){Circular Buffer}}
\put(12.5,0){\framebox(50,15){\shortstack{DRAM I/O \\ and EDC}}}
\put(36,35){\line(0,-1){20}}
\put(19,15){\line(0,1){5}}
\put(19,30){\line(0,1){5}}
\put(55,15){\line(0,1){5}}
\put(55,30){\line(0,1){5}}

\end{picture}

\caption{The main components of the NGA.}
\label{fig:nga}
\end{figure}
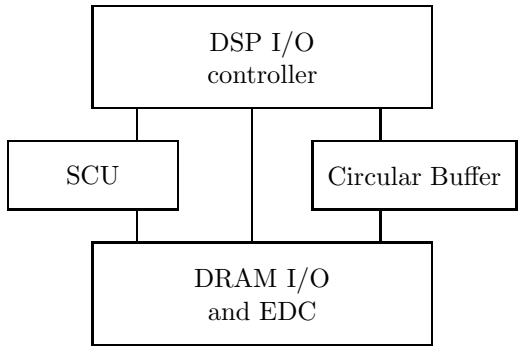

The Columbia 0.8 Teraflops computer is a collection of $16,384$ {\em nodes}
connected in a $16^{3} \times 4$ four-dimensional, serial network. The
main elements of the node (Fig. \ref{fig:node}) are:

\begin{itemize}
\item Digital Signal Processor (DSP) capable of performing a floating
point multiplication and  an accumulation 
in a single 25MHz cycle.
We are currently using Texas Instruments TMS320C30 50MHz parts.
For more information on DSP applications see \cite{DSP}.

\item Dynamic~Random~Access~Memory (DRAM). We are supporting either 
$256k \times 16$ or $512k \times 8$ parts from various vendors. These
memory devices are connected to a 39-bit data bus (32-bit data word 
with 7 bits for Error Detection and Correction).

\item The cornerstone of our design, the NGA (Node Gate
Array), is a customized gate array (Fig. \ref{fig:nga}).
The NGA provides a controller for the serial communication protocol, arbitrates 
memory accesses, handles recovery from errors in DRAM and serial wires, and
includes a buffer that balances data transfers between the DSP and
DRAM.
\end{itemize}

  \pagenumbering{arabic}
  \addtocounter{page}{1}

The opportunity to include almost all I/O controllers and global signal
handlers in one VLSI chip significantly simplified the configuration of
the node  and at the same time added a new level of flexibility
to the hardware. In designing the NGA, we tried to
shift as much of the  low level functionality as possible to the hardware
thus allowing the programmer to concentrate more on the implementation
of the physics algorithms. At the same time we are leaving enough hooks
for a sophisticated user concerned with getting maximum efficiency.

The design of a truly parallel computer requires a well balanced
communication scheme with the throughput matching the speed of the CPU.
The optimal communications vs. computations ratio depends on
the algorithms to be run on the computer. Our task was to produce
a Lattice QCD machine able to run full QCD calculations efficiently. Our
benchmark algorithm for determining balance with the
serial communications and memory accesses is a single conjugate
gradient (CG) update from a staggered fermion evolution. This is more or less
an obvious choice for the full QCD calculation since matrix inversions take
up most of the CPU time. We also used code for  Wilson fermions which,
though less communications intensive, provides a useful tool for estimating
overall performance.
A serial communication scheme with
transfers going in four "space--time" directions 
does not create substantial overhead for vector--matrix multiplications
if run at a speed exceeding 17MHz. 
On the other hand,
global dot products that require the sum of the spinors on all nodes would take
up to 70\% of the CPU time if serial transfers were run at 25MHz. Our SCU 
(Serial Communication Unit) thus has been designed to perform  
``on-the-fly'' add, max, and broadcast operations between
the data residing on a particular node and the data coming from the 
adjacent nodes. This mode significantly reduces the latency of
transferring information across the computer.

Another aspect of achieving balance among different components of the node
is managing slow DRAM. This is done by introducing a pipeline stage
in the path between the DSP and the memory. A 32-word Circular Buffer
can be instructed to prefetch a certain number of words from DRAM which
can be read by the DSP in zero wait state mode later on. The Circular
Buffer has an embedded protection against accesses to invalid data
locations unless the programmer specifically turns this protection off.
The size of the Circular Buffer comfortably accommodates 
all eighteen elements of a $SU(3)$ color matrix. 

\section{MODERN DESIGN TOOLS}

The NGA incorporates almost all the nontrivial functionality that makes a
collection of CPU's and memory devices a parallel computer. Please,
refer to \cite{Mawh} or \cite{NHC} for the description
of the upper level design blocks of our project. In this section we will 
concentrate on the implementation of the ideas described in the the
previous section. 

Not long ago, at the final stages of the design process, one had to
produce  schematic drawings, then solder available logic
elements and more sophisticated standardized medium integration
components on printed circuit boards, and test the result using a 
logical analyzer. The design process was time consuming and
hardly feasible outside large companies or research centers. 
A major advance  occurred a few years ago with the advent of {\em hardware
description languages}, such as VHDL and Verilog, that allow a designer
to formalize some higher level concepts and provide a high degree of
automation in transforming this description into an actual schematic
accepted by VLSI manufacturers. 

The basic entities that VHDL (our language of choice) describes are
signals and their sequential and concurrent assignments. One can write 
boolean style logic equations connecting incoming signals such as the DSP
address bus with outgoing signals such as serial wires constituting the
communication network or controls for peripherial devices. {\em Finite state
machines}, that react to the incoming signals according to the {\em
state} the machine is in, are naturally implemented as {\em clocked
processes} where the state is represented by an internal signal
that is allowed to change value only at a specific edge of the clock.
The following is an example of a two-bit counter:

\begin{flushleft}
{\small\tt 
 \hspace{5mm} wait until prising(clock); \\
 \hspace{5mm} count0 $<$= not count0; \\
 \hspace{5mm} count1 $<$= count1 xor count0; 
}
\end{flushleft}

Once the desired logical equations are written, the design can be {\em
simulated} as a black box which reacts to the external stimuli
represented by time patterns for each incoming signal. We also use models
for the 
standard components such as DSPs and DRAMs available from the Logic Modeling
Corporation which allow us to 
check the logical consistency of a design by running a piece
of CG code on a model of the DSP which communicates with our model of the
NGA which, in turn, reads and writes to a model of the memory.

The next important step is {\em synthesis} of the  VHDL source code; the
output of which is the desired schematic. This process is similar to
compiling an ordinary program. The gate array manufacturer supplies
a library of components which represent basic constructs in VHDL. The
{\em synthesizer} parses the code and substitutes these constructs with
the elements of the library, properly connecting them to each other.
After the initial run, the synthesizer proceeds to the iterative
optimization task by looking at larger clusters of components and trying
to reduce them to more compact ones following the list of constraints such
as speed, size, or load requirements set by the user. The development
of efficient synthesis tools is very much in progress now and the
current ones are not perfect. Some human intervention is still
required to produce sound results. For more information on this aspect
of design see \cite{VHDL}.

The final step in the design is the simulation which includes
correct propagation delays supplied by the vendor. At this stage one can
identify parts of the design that are too slow to keep up with the
desired speed of the computer and optimize them either by
rearranging the VHDL code or by applying more severe constraints to the
synthesizer. There are a number of tools available for this: one can
list the asynchronous paths with delays larger than a set number,
include such signals in a watch list from the simulator, look at
the diagnostic output of the synthesizer, and so on. At the same time
one can run large pieces of assembly code on the model of the DSP, checking
the performance of physics code on a cycle~by~cycle basis and revealing
obscure bugs that might otherwise show up only after thousands of cycles of
simulation. The performance figures for the Columbia 0.8 Teraflops
project are based on these simulations and presented in reference \cite{Mawh}.

\end{document}